\documentclass[aps,prl,superscriptaddress,amsmath,amssymb,reprint,floatfix,showpacs]{revtex4-1}

\usepackage{graphicx}
\usepackage{bm}

\begin{document}

\title{Growth and phase velocity of
self-modulated beam-driven plasma waves}

\author{C. B. Schroeder}
\author{C. Benedetti}
\author{E. Esarey}
\affiliation{Lawrence Berkeley National Laboratory, Berkeley,
California 94720, USA}
 
\author{F. J. Gr\"{u}ner} 
\affiliation{Universit\"{a}t Hamburg, Luruper Chaussee 149, 22761 Hamburg, Germany}
 
\author{W. P. Leemans}
\affiliation{Lawrence Berkeley National Laboratory, Berkeley,
California 94720, USA}

\date{Submitted to {\it Physical Review Letters} 19 July 2011}

\begin{abstract}
A long, relativistic charged particle beam propagating in a plasma is
subject to the self-modulation instability.  This instability is
analyzed and the growth rate is calculated, including the phase
relation.  The phase velocity of the accelerating field is shown to be
significantly less than the drive beam velocity.  These results
indicate that the energy gain of a plasma accelerator driven by a
self-modulated beam will be severely limited by dephasing.  In the
long-beam, strongly-coupled regime, dephasing is reached in less than
four e-foldings, independent of beam-plasma parameters.
\end{abstract}


\maketitle

Plasma-based accelerators have attracted considerable attention owing
to the ultrahigh field gradients sustainable in an electron plasma
wave, enabling compact accelerators.  The electric field amplitude of
the electron plasma wave (space-charge oscillation) is on the order of
$E_0 = cm_e\omega_p/e$, or $E_0 [{\rm V/m}] \simeq 96 \sqrt{n_0 [{\rm
cm}^{-3}]}$, where $\omega_p = (4\pi n_0 e^2/m_e)^{1/2}$ is the
electron plasma frequency, $n_0$ is the ambient electron number
density, $m_e$ and $e$ are the electron rest mass and charge,
respectively, and $c$ is the speed of light in vacuum.  This field
amplitude can be several orders of magnitude greater than conventional
accelerators.  Electron plasma waves with relativistic phase
velocities may be excited by the nonlinear ponderomotive force of an
intense laser \cite{Esarey09} or the space-charge force of a charged
particle beam, i.e., a plasma wakefield accelerator (PWFA)
\cite{Chen85,Rosenzweig88}.  In 2006, high quality 1~GeV electron
beams were produced using using 40~TW laser pulses in cm-scale plasmas
\cite{Leemans06b}.  In 2007, a 42 GeV electron beam in a meter-long
plasma was used to double the energy of a small fraction of electrons
on the beam tail by the plasma wave excited by the beam head
\cite{Blumenfeld07}.
These experimental successes have resulted in further interest in the
development of plasma-based acceleration as a basis for future
linear colliders  \cite{Seryi09,Schroeder10b}.

It has recently been proposed to drive a plasma accelerator with a
highly relativistic proton beam, such as those available at CERN
(European Organization for Nuclear Research)
\cite{Caldwell09,Lotov10}.  In general, exciting plasma waves requires
a drive beam density profile with frequency components at the plasma
frequency, i.e., a beam density longitudinal scale length on the order
of the plasma wavelength $\lambda_p= 2\pi/k_p = 2\pi c/\omega_p$, or
$\lambda_p[\mu{\rm m}] = 3.3\times 10^{10}/\sqrt{n[{\rm cm}^{-3}]}$.
Compact, high-gradient accelerators require high plasma density, and
therefore require short drive beams, e.g., $\lambda_{p}\sim 100\ \mu$m
for $n_{0}\sim 10^{17}$ cm$^{-3}$.  Generating short proton beams (or
proton beams with spatial structure at $\lambda_{p}$) is challenging,
and it has been proposed to rely on a beam-plasma instability to
modulate the beam at $\lambda_{p}$, driving a large amplitude plasma
wave \cite{Kumar10}.  The self-modulation of the beam occurs through
coupling of the transverse wakefield with the beam radius evolution.
Periodic regions of focusing and defocusing modulate the beam density
at $\lambda_{p}$, driving a larger plasma density modulation that
further focuses the beam periodically.  This is somewhat similar to the
self-modulation instability that occurs for long laser pulses
\cite{Esarey94}.  For beams long compared to $\lambda_{p}$, where
self-modulation occurs, the instability is enabled by the drive beam
dynamics, and therefore the wakefield properties will be strongly
affected by the drive beam dynamics.

An important quantity characterizing the performance of a plasma
accelerator is the phase velocity $v_{p}$ of the plasma wave.  For
$v_{p}<c$, a highly relativistic electron will outrun the
plasma wave and phase slip from the accelerating to the decelerating
phase region of the plasma wave.  This limits the electron energy gain
to $\Delta W\sim \gamma_{p}^2 (E_{z}/E_{0}) m_{e}c^{2}$ after
acceleration over a dephasing length $L_{d}\sim
\gamma_{p}^{2}\lambda_{p}$, where $E_{z}$ is the electric field
amplitude of the plasma wave and
$\gamma_p=(1-v_{p}^{2}/c^{2})^{-1/2}$.  For a plasma accelerator
driven by a short ($<\lambda_{p}$) intense laser pulse, $v_{p}$ can be
relatively low ($\gamma_{p}\sim 10-100$) and dephasing can limit the
energy gain \cite{Schroeder11}.  For a PWFA driven by a short
($<\lambda_{p}$) highly-relativistic beam, $v_{p}$ can be
sufficiently high so that dephasing is not an issue.

In this Letter we calculate the self-modulation of particle beams in
plasma, including the properties of the excited plasma wave.  In
particular, we show that the phase velocity of the plasma wave excited
by self-modulation is greatly reduced from the velocity of the drive
beam.  The phase velocity is determined by the growth of the
instability and the beam-plasma dynamics.  A similar effect occurs in
self-modulated laser-driven plasma waves \cite{Andreev96,Leemans96b}.
Analytic solutions for the growth rate and phase velocity in the long
beam regime are derived and compared to numerical solutions of the
envelope equation for the particle beam.  Owing to the low
phase velocity of the plasma wave, the maximum energy gain in such a
self-modulated beam-driven accelerator will be severely limited by
dephasing.

The wake generated by a relativistic particle beam driver
moving through a plasma can be calculated using the cold plasma fluid
and Maxwell equations.  Here we consider a drive beam consisting of
particles with charge $\mp e$ and mass $M_b$.  In the linear wake 
regime, the normalized density
perturbation $\delta n/n_0 = (n - n_0)/n_0< 1$ driven by a beam with
density $n_b< n_0$ is
\begin{equation}
\left( \partial^2_\zeta + k_p^2 \right) \delta n/n_0 =  
\mp k_p^2 n_b/n_0 ,
\label{eq:n}
\end{equation}
where the $\mp$ corresponds to a negatively/positively charged
particle beam.  A highly relativistic beam is assumed with Lorentz
factor $\gamma = (1-\beta_b^2)^{-1/2} \gg 1$, and the quasi-static
approximation is taken such that the plasma fluid quantities are
functions of the co-moving variable $\zeta = z-\beta_b t$.  The
beam-driven longitudinal electric field $E_{z}$ and transverse fields 
$E_{r}$ and $B_{\theta}$ are \cite{Keinigs87}
 \begin{gather}
 \left( \nabla_\perp^2 - k_p^2 \right) E_z/E_0 = -k_p \partial_\zeta 
 \delta n/n_0 ,
 \label{eq:Ez}
 \\
 \left( \nabla_\perp^2 - k_p^2 \right) (E_r -B_\theta)/E_0 = 
 -k_p \partial_r 
 \delta n/n_0
\label{eq:Wr}
.
\end{gather}
The transverse beam-driven wakefield Eq.~\eqref{eq:Wr} is coupled to
the envelope equation for the beam \cite{Reiser08}
\begin{equation}
\frac{d^2 R}{dz^2} - \frac{\epsilon_n^2}{4\gamma^2R^3} = 
\mp\frac{ 1}{\gamma R}
\frac{m_e}{M_b}
\left\langle k_p r \left( E_r - B_\theta \right)/E_0
\right\rangle
,
\label{eq:env-1}
\end{equation}
where $R = \langle r^2 \rangle^{1/2}$ is the rms beam size,
$\epsilon_n = \gamma [\langle r^2 \rangle \langle ( {d r}/{dz}
)^2\rangle - \langle r {d r}/{dz} \rangle^2 ]^{1/2}/2$ is the
normalized transverse emittance in cylindrical geometry, and the
brackets indicate an average over the transverse beam distribution.

For simplicity, in the following we consider a beam with a flat-top
radial profile, $n_b= [n_{b0}r_{b0}^2/r_b^2] f(\zeta) \Theta (r-r_b)$,
where $f$ is the normalized longitudinal profile, $\Theta$ is the
Heaviside function, $r_b (\zeta, z)$ is the beam radius, and $r_{b0} =
r_b (\zeta, z=0)$ is the initial beam radius.  For a flat-top radial
profile, Eqs.~\eqref{eq:n} and \eqref{eq:Wr} have the solution
\begin{multline}
(E_r-B_\theta)/E_0 = \pm ({n_{b0}}/{n_0})k_p^2r_{b0}^2I_1 (k_p r)
\\
\times 
\int_{\infty}^{\zeta}  d\zeta'\sin [k_p (\zeta-\zeta' )] 
f(\zeta') K_1 (k_pr_b(\zeta')) /r_b(\zeta') 
,
\label{eq:Wr-flat}
\end{multline}
for $r\leq r_b$, assuming the initial radius $r_{b0}$ is independent
of $\zeta$.  Here $I_m$ and $K_m$ are the modified Bessel functions.
Using Eqs.~\eqref{eq:env-1} and \eqref{eq:Wr-flat}, the envelope
equation for the beam radius $r_b(\zeta, z) = \sqrt{2} R$ at any slice
$\zeta$ is
\begin{multline}
\frac{d^2 r_b}{dz^2} - \frac{\epsilon_n^2}{\gamma^2r_b^3} 
 =  
-\frac{4 k_b^2 r_{b0}^2 I_2(k_pr_b)}{\gamma r_b} 
\\ \times 
\int^{\zeta}_{\infty} d\zeta'\sin [k_p (\zeta-\zeta' )]
f(\zeta') K_1 (k_pr_b(\zeta')) /r_b(\zeta') , 
\label{eq:env-flat}
\end{multline}
where $k_b^2 = 4 \pi n_{b0} e^2 /M_b$ is plasma wavenumber of the
beam.  Equation~\eqref{eq:env-flat} describes the coupled beam
evolution and wakefield excitation.

Consider a long beam compared to the plasma wavelength, where the
variation in the longitudinal profile may be neglected $f(\zeta)\simeq
1$, propagating in a plasma with a perturbation at the plasma
wavelength.  In the following we will decompose the plasma and beam
quantities such that $Q = Q_0+Q_1$, where the `0' subscripts indicate
the long beam solution and the `1' subscripts indicate the
perturbation.  The initial perturbation or instability seed may be due
to excitation of a plasma wave from the head of the beam, fluctuations
in the beam or plasma, or from a plasma wave externally excited (e.g.,
by a short-pulse laser).
Using Eqs.~\eqref{eq:n}--\eqref{eq:env-1}, the unperturbed long beam
solution is $(\delta n)_0 = -n_b$, $(E_z)_0 =0$, $(E_r-B_\theta)_0
= \pm E_0 ({n_{b0}}/{n_0}) k_p r_0 K_1 (k_pr_0) I_1 (k_p r)
(r_{b0}/r_0)^2$ for $r\leq r_0$, and the beam radius evolves as
\begin{equation}
\frac{d^2 r_0}{dz^2} - \frac{\epsilon_n^2}{\gamma^2r_0^3} + 
\frac{4k_b^2 r_{b0}^2}{\gamma k_p r_0^2} 
  K_1 (k_pr_0)I_2(k_pr_0) =0 .
\end{equation}
We will assume that the beam is initially in the long beam equilibrium
$r_{b0} = r_0 = r_{\rm eq}$, such that $d^2r_0/dz^2=0$, and $r_{\rm
eq}$ is given by ${\epsilon_n^2k_p} = {4 \gamma k_b^2}r_{\rm eq}^3 K_1
(k_pr_{\rm eq})I_2(k_pr_{\rm eq}) $.  For a narrow beam, $k_pr_{\rm
eq} \ll 1$, the equilibrium beam radius is $r_{\rm eq} = [ {2
\epsilon_n^2}/{ k_b^2 \gamma } ]^{1/4}$.

Assuming a small perturbation about this equilibrium, $r_b = r_0 +
r_1$ with $\vert r_1 /r_0\vert \ll 1$, and expanding
Eq.~\eqref{eq:env-flat} yields the evolution of the beam radius
perturbation
\begin{equation}
\left( \frac{d^2 }{d\hat z^2} + 4 \kappa^2 \right) {r_1} 
= 
2 \nu
\int^{\zeta}_\infty  d\hat\zeta'\sin (\hat \zeta - \hat \zeta' )
 r_1 (\hat \zeta') ,
\label{eq:lin}
\end{equation}
with the constants
\begin{equation}
\kappa^2 =
2 K_1 (k_pr_0 ) 
\left[4 \frac{I_2(k_p r_0)}{k_p r_0} + {I_3(k_p r_0)} \right] ,
\label{eq:kappa}
\end{equation}
and $\nu ={4 I_2(k_pr_0)} K_2 (k_pr_0) $, and the normalized variables
$\hat \zeta = k_p \zeta$ and $\hat z = k_b z /(2\gamma)^{1/2}$.  In
the limit of a narrow beam $k_p r_0 \ll 1$, $\nu \simeq 1 -
(k_pr_0)^2/6 $, and $\kappa^2 \simeq 1 + {(k_pr_0)^2 } [ C_\gamma -1/4
+ \ln (k_pr_0/2)]/2 $, where $C_\gamma \simeq 0.577$ is the
EulerÐ-Mascheroni constant.
Equation \eqref{eq:lin} may be analyzed
in several regimes.  The most relevant regime for plasma accelerators
based on self-modulated drive beams is the strongly-coupled (or
long-beam, early-time) regime valid for $\hat \zeta \gg \hat z$.

Applying the linear plasma wave operator to Eq.~\eqref{eq:lin} yields
\begin{equation}
( \partial^2_{\hat \zeta} + 1 ) 
( \partial^2_{\hat z} +4  \kappa^2 ) {r_1} 
= 2 \nu   r_1 .
 \label{eq:lin2}
\end{equation}
Consider a slowly varying envelope, such that $r_1 = \hat r \exp (ik_p
\zeta )/2 + \textrm{c.c.}$ with $\vert \partial_{\zeta} \hat r \vert
\ll \vert k_p \hat r \vert$, and assume the strongly-coupled regime
where the growth length of the instability is short compared to
$\gamma^{1/2}k_b^{-1}$, such that $ \vert \partial_{\hat z} \hat r
\vert \gg 2 \kappa \vert \hat r \vert $.  In this regime
Eq.~\eqref{eq:lin2} becomes
\begin{equation}
\left( \partial_{\hat \zeta}  \partial^2_{\hat z} + i \nu \right)
 {\hat r} = 0 ,
\label{eq:smi}
\end{equation}
which describes the evolution of the slowly varying amplitude of the
beam radius perturbation and may be solved using standard Laplace
transform techniques.  With the initial conditions $\hat r (z,
\zeta=0) = \delta r \Theta(z)$, $\hat r (z=0, \zeta) = \delta r$, and
$\partial_z \hat r (z=0, \zeta) = 0$, the solution to
Eq.~\eqref{eq:smi} can be expressed as
\begin{equation}
\hat r  /\delta r =  \sum_{n=0}^{\infty} 
\frac{(i\nu\vert\hat\zeta\vert \hat z^2)^n}{n!(2n)!}
= {_0\mathsf{F}_2} \bm{(}; \{ 1/2, 1\} ; i \nu \vert \hat \zeta\vert 
\hat z^2/4\bm{) } , 
\label{eq:hypgeo}
\end{equation}
where $_q\mathsf{F}_p$ is the generalized hypergeometric function.
The solution to Eq.~~\eqref{eq:smi} may also be evaluated
asymptotically and has the form
\begin{equation}
r_1 
= \delta r \frac{ 3^{1/4}}{(8\pi)^{1/2}} 
N^{-1/2}{e^{N}}
\cos \left( k_p \zeta + N/\sqrt{3} - \pi/12 \right) ,
\label{eq:hatr}
\end{equation}
where the number of e-foldings is 
\begin{equation}
N = \frac{3^{3/2}}{4} 
\left( \nu \frac{n_b m_e }{n_0 M_b\gamma} 
k_p^3\vert \zeta \vert z^2\right)^{1/3} 
.
\label{efold}
\end{equation}
Note that growth Eq.~\eqref{efold} [and the beam envelope equation,
Eq.~\eqref{eq:env-flat}] differ from that found in
Ref.~\cite{Kumar10}.

\begin{figure}[t!]
    \begin{center}
 \includegraphics[scale=1]{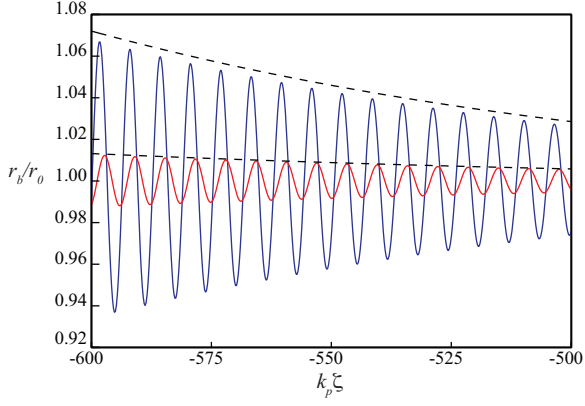}
    \end{center}
\caption{\label{fig:r} (Color online) Beam radius modulation $r_b/r_0$
vs $k_p \zeta$ with beam-plasma parameters $n_b/n_0=0.008$, $\gamma=
107$, and $k_pr_0=1$ (and $r_{b0} = r_0 = r_{\rm eq}$), obtained from
numerical solution of Eq.~\eqref{eq:env-flat}, at $k_pz = 8000$ (red
curve) and $k_pz = 9500$ (blue curve).  Dashed curves are
the envelope of the asymptotic linear solution Eq.~\eqref{eq:hatr}.}
   \end{figure}

Figure \ref{fig:r} shows the beam radius modulation $r_b/r_0 = 1 +
r_1$ versus $k_p\zeta$, after propagating $k_pz = 8000$ (red curve)
and $k_pz = 9500$ (blue curve), obtained from numerical solution of
Eq.~\eqref{eq:env-flat} for a beam initially in equilibrium $r_{b0} =
r_0 = r_{\rm eq}$ with beam-plasma parameters $n_b/n_0=0.008$, $\gamma=
107$, and $k_pr_0=1$.  The dashed curves are the envelope of the
linear asymptotic solution Eq.~\eqref{eq:hatr}.  Figure \ref{fig:r}
shows the growth versus distance behind the head of the beam
(at $k_p\zeta = 0$) and versus propagation distance.  Also shown is the
shift in phase of the modulation versus propagation distance,
resulting in a reduced phase velocity, as discussed below.

The above solution Eq.~\eqref{eq:hypgeo} assumed $\vert k_p \hat r
\vert \gg \vert \partial_\zeta \hat r \vert$, or $1 \gg \vert k_p^{-1}
(\partial_\zeta N) \vert$.  This condition may be expressed as
\begin{equation}
1 \gg \frac{3^{3/2}}{2^6} \left( \frac{k_b^2}{k_p^2}
\frac{\nu}{\gamma} \right) \left( \frac{z}{\vert \zeta 
\vert} \right)^{2}
,
\label{eq:val}
\end{equation}
or $\hat \zeta \gg \hat z$, which will be satisfied for long beams
sufficiently early in the beam propagation.  It was also assumed that
$\vert \partial_{\hat z} \hat r \vert \gg 2 \kappa \vert \hat r\vert$,
which is satisfied provided Eq.~\eqref{eq:val} is satisfied.  The
above analysis is also based on linear theory, and nonlinear effects
(i.e., when $r_1 \sim r_0$ or $E_z \sim E_0$) may saturate the
instability.

The beam radius perturbation $r_1 = \hat r \exp (ik_p \zeta)/2 + {\rm
c.c.}$ modulates beam density $n_b \simeq n_b (r_0) (1 - 2 r_1/r_0)$.
This beam density modulation drives a modulation in the electron
plasma density $\hat n \exp (ik_p \zeta)/2 + {\rm c.c.}$, via
Eq.~\eqref{eq:n}, i.e., $\partial_\zeta \hat n \simeq \mp i k_p
n_b(r_0) \hat r/r_0$.  The plasma density modulation drives the
accelerating wakefield $E_z/E_0 = \hat E_z \exp(ik_p\zeta)/2 + {\rm
c.c.}$, via Eq.~\eqref{eq:Ez}, i.e., $(\nabla_\perp^2 -k_p^2 )
\partial_\zeta \hat E_z = \mp k_p^3 [n_b(r_0)/n_0] \hat r/r_0$.  For
the same initial conditions as above, the series solution for the
accelerating wakefield in the long-beam regime is
\begin{equation}
\hat E_z  = \mp H_R(r,r_0) \frac{n_{b0}}{n_0} \frac{\delta r}{r_0} 
\vert\hat \zeta\vert\sum_{n=0}^{\infty} \frac{(i\nu\vert\hat\zeta\vert
\hat z^2)^n}{(n+1)!(2n)!} ,
\label{eq:ez-series}
\end{equation}
and the sum may be expressed as the hypergeometric function
${_0\mathsf{F}_2} \bm{(}; \{ 1/2, 2 \} ; i \nu \vert \hat\zeta\vert
\hat z^2/4\bm{) }$.  Here $H_R(r,r_0) = 1 - k_pr_0 K_{1}(k_pr_0 ) I_0
(k_pr )$ for $r\leq r_0$ and $k_pr_0 I_{1}(k_pr_0 ) K_0 (k_pr )$ for
$r> r_0$.  In the asymptotic limit, $ E_z / E_z(z=0) \simeq 3^{7/4}(32
\pi)^{-1/2} N^{-3/2}\exp (N) \cos (\psi)$, where the number of
e-foldings of growth of the accelerating wake is given by
Eq.~\eqref{efold} and the phase is
\begin{equation}
\psi = k_p \zeta - \frac{\pi}{4}  + \frac{3}{4} 
\left( \nu \frac{k_b^2 k_p}{\gamma} \vert \zeta \vert z^2\right)^{1/3} .
\label{eq:phase}
\end{equation}

The phase velocity of the accelerating wake is given by $\beta_p = -
\partial_t \psi / \partial_z \psi = \partial_\zeta \psi /
(\partial_\zeta + \partial_z)\psi \simeq 1 -
\partial_z\psi/\partial_\zeta \psi$.  In this regime, i.e., satisfying
Eq.~\eqref{eq:val}, $ \beta_p \simeq 1 - k_p^{-1} \partial_z\psi $.
Using the phase Eq.~\eqref{eq:phase}, the phase velocity is $\beta_p =
1 - ({2}/{3^{3/2}}) (N/{k_p z})$.
The phase velocity of the self-modulated beam-driven wakefield is less
than the beam velocity $\beta_b\simeq 1$, varies along the beam
$\zeta$ and during propagation $z$.  Asymptotically, the Lorentz
factor of the phase velocity is
\begin{equation}
\gamma_p = \left( \frac{\gamma n_0 M_b}{\nu n_{b0} m_e} 
\frac{z}{\vert \zeta \vert }  \right)^{1/6} 
\label{eq:phasevel}
\end{equation}
in the strongly-coupled, long-beam regime.  Note that, behind the
modulated beam the phase velocity is given by Eq.~\eqref{eq:phasevel}
with $\vert \zeta \vert = L_b$, where $L_b$ is the bunch length.

  \begin{figure}[t!]
   \begin{center}
\includegraphics[scale=1]{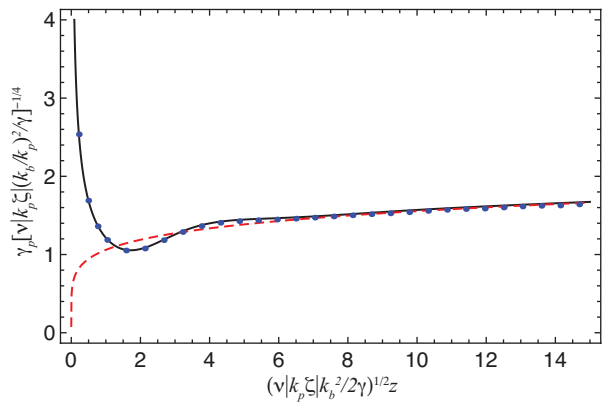}
   \end{center}
   \caption{\label{fig:pv}
(Color online) Normalized phase velocity of accelerating wakefield
$\gamma_p [\nu (k_b/k_p)^2 \vert \hat \zeta \vert / \gamma ]^{-1/4}$
vs.\ normalized propagation distance $(\nu \vert \hat \zeta
\vert)^{1/2} \hat z$ in the long-beam regime: solid (black) curve is
the series solution Eq.~\eqref{eq:ez-series}, dashed (red) curve is
the asymptotic solution Eq.~\eqref{eq:phasevel}, and (blue) dots are
from the numerical solution of the envelope equation
Eq.~\eqref{eq:env-flat}.}
  \end{figure}

Figure \ref{fig:pv} shows the normalized Lorentz factor of the phase
velocity of the accelerating wakefield $\gamma_p [\nu (k_b/k_p)^2
\vert \hat \zeta \vert / \gamma ]^{-1/4}$ versus normalized
propagation distance $(\nu \vert \hat \zeta \vert)^{1/2} \hat z$.  The
solid curve in Fig.~\ref{fig:pv} is obtained from the series solution
Eq.~\eqref{eq:ez-series}, $\beta_p = 1 - k_p^{-1}\partial_z [\arctan
(\Im \hat E_z / \Re \hat E_z)]$, the dashed curve is the asymptotic
solution Eq.~\eqref{eq:phasevel}, and the dots are from the numerical
solution (with the parameters $\gamma=107$, $n_{b0}/n_0 = 0.008$, $k_p
r_0= 1$, and $k_p L_b = 600$) of the envelope equation
Eq.~\eqref{eq:env-flat}.  
Figure~\ref{fig:pv} indicates that there is a minimum phase velocity.
The minimum phase velocity can be estimated by using the series
solution Eq.~\eqref{eq:ez-series}.  The minimum phase velocity occurs
at
$(\nu \vert \hat \zeta \vert)^{1/2} \hat z \simeq 1.72$,
with 
\begin{equation}
\gamma_{\rm min}
\simeq 1.06 \left(  \frac{\gamma n_0 M_b}{\nu n_{b0} 
m_e  k_p\vert \zeta 
\vert } \right)^{1/4}
.
\end{equation}
As shown in Fig.~\ref{fig:pv}, after reaching $\gamma_{\rm min}$, the
phase velocity grows slowly as the beam propagates $\gamma_p \propto
z^{1/6}$ [cf.~Eq.~\eqref{eq:phasevel}].  For example, consider a
wake driven by a 100~GeV proton beam ($\gamma = 107$), with $
r_0 = 180~\mu$m, $L_b = 10$~cm, and $10^{11}$ particles.
Operating at $n_0=10^{15}~\textrm{cm}^{-3}$, corresponds to
$n_{b0}/n_0=0.008$, $E_0=3$~GV/m, $k_p r_0 = 1.0$, $k_p L_b = 600$,
and $\nu= 0.88$.  For this example, $\gamma_{\rm min} \simeq 15$
behind the drive beam after $z\simeq 8.5$~cm (i.e., $\approx
500~\lambda_p$) of propagation.

With the phase velocity of the self-modulated wake determined, the
dephasing length may be calculated.  For a linear wake, the dephasing
length is the propagation distance required for an ultra-relativistic
particle $\beta_b \simeq 1$ to slip $\lambda_p/4$ (or a wake phase of
$\pi/2$) with respect to the plasma wave.  Assuming the phase velocity
is well-approximated by the asymptotic solution in the
strongly-coupled regime Eq.~\eqref{eq:phasevel}, the dephasing length
is $L_d =( {2\pi}/{3} )^{3/2} ( \nu k_b^2k_p\vert \zeta_i \vert
/\gamma)^{-1/2}$.  Including the early time response via
Eq.~\eqref{eq:ez-series}, the dephasing length is
\begin{equation}
L_d \simeq 4.9 \left( \nu k_b^2 k_p
\vert \zeta_i\vert / \gamma \right)^{-1/2} ,
\label{eq:Ld}
\end{equation}
where $\zeta_i$ is the injection position of the witness bunch (e.g.,
initially at a peak of the accelerating field).  For a witness bunch
injected behind the drive beam, $\zeta_i = L_b$.  This reduced
dephasing length will greatly limit the energy gain of a witness
electron beam trailing the drive bunch.  For example the number of
e-foldings of the self-modulational instability that have occurred at
the dephasing length Eq.~\eqref{eq:Ld} is $ N ( z = L_d, \zeta_i)
\simeq 3.8$.  Note that the number of e-foldings at a dephasing length
$N ( z = L_d) $ is independent of injection location and the
beam-plasma parameters.
Improved efficiency may be possible by tapering the plasma density,
i.e., increasing the background plasma density to reduce the plasma
wavelength, thereby increasing the phase velocity \cite{Katsouleas86},
although variation of the density may affect the instability growth.
Alternatively the accelerator may use a staged approach, where a long
plasma region self-modulates the drive beam, followed by a second
stage where a witness bunch would be injected following the modulated
drive beam.  Such a two-staged approach could potentially be limited
by the hose (or transverse two-stream) instability \cite{Whittum93},
which grows in the long beam limit with a comparable growth rate $\sim
N$.  This implies that to drive large amplitude accelerating fields
via the self-modulational instability without hosing requires strongly
seeding the instability.  One possibility is to use a beam with a fast
rise in the current profile \cite{Kumar10}.  Another possibility to
seed the modulation is via an intense short-pulse laser.

The long-beam, early-time regime described above will be valid for
$\hat z \ll \hat \zeta$.  After sufficiently long propagation
distances, or for sufficiently short beams, the instability may enter
a weakly-coupled regime where the instability growth length is long
compared to $\gamma^{1/2}/k_b$.  The instability will transition to
the weakly-coupled regime after a propagation distance approximately
$\hat z \sim \hat \zeta$, or, using Eq.~\eqref{efold}, after
approximately $N \sim k_p \zeta$.  For long beams $k_p
\zeta \gg 1$, nonlinear effects will typically appear before the
instability enters the weakly-coupled regime.

In this Letter we have calculated the beam self-modulation instability
growth rate, in the long-beam regime, including the phase dependence.
The phase velocity of the accelerating wakefield was calculated and
shown to be significantly less than the drive beam velocity.  The
dephasing length was calculated, and, in the strongly-coupled regime,
a witness beam will reach dephasing in less than four e-foldings,
independent of beam-plasma parameters.  This indicates that the energy
gain in a plasma accelerator driven by a self-modulated PWFA will be
limited by dephasing.

This work was supported by the Director, Office of Science, Office of
High Energy Physics, of the U.S.~Department of Energy under Contract
No.~DE-AC02-05CH11231.

\bibliographystyle{popsty2}

\end{document}